\documentclass[conference,compsoc]{IEEEtran}
\ifCLASSOPTIONcompsoc
  \usepackage[nocompress]{cite}
\else
  \usepackage{cite}
\fi
\usepackage{epsfig,endnotes}
\usepackage{textcomp,amssymb}
\usepackage{setspace}
\usepackage{amsfonts}
\usepackage{latexsym,fancyhdr,url}
\usepackage{enumerate}
\usepackage{enumitem}
\usepackage{amsmath}
\usepackage[ruled,vlined,linesnumbered]{algorithm2e}
\usepackage{algorithm2e}
\usepackage{algpseudocode}
\usepackage{booktabs}
\usepackage{makecell}
\usepackage{graphics}
\usepackage{xparse} % argument parsing -- \edist
\usepackage{xspace}
\usepackage{multirow}
\usepackage{csvsimple}
\usepackage{balance}
\usepackage{makecell}
\usepackage{pifont}
\usepackage{enumitem}
\usepackage{colortbl}
\usepackage{placeins}
\usepackage{fancyhdr}
\usepackage{listings}
\usepackage{caption}
\usepackage[most]{tcolorbox}
\usepackage{wasysym}

\usepackage{titlesec}
\titleformat{\paragraph}[runin]
  {\normalfont\normalsize\bfseries}{\theparagraph}{1em}{}
\titlespacing*{\paragraph}
  {0pt}{3.25ex plus 1ex minus .2ex}{1em}

\usepackage{hyperref}

\definecolor{refcolor}{RGB}{190,0,0}
\definecolor{urlcolor}{RGB}{0,0,0}

\hypersetup{
    colorlinks=true,
    linkcolor=refcolor,
    citecolor=refcolor, 
    urlcolor=urlcolor,
}

\usepackage{listings}
\usepackage{xcolor}
\definecolor{keywordcolor}{HTML}{b00040}
\definecolor{keywordcolor2}{HTML}{008000}
\definecolor{typecolor}{rgb}{0.17,0.57,0.68}
\definecolor{functioncolor}{HTML}{1818ff}
\definecolor{green}{rgb}{0,0.6,0}
\definecolor{red}{rgb}{0.6,0,0}
\definecolor{gray}{rgb}{0.5,0.5,0.5}

\lstdefinestyle{mystyle}{
    backgroundcolor=\color{white},   
    commentstyle=\color{green},
    keywordstyle=\color{keywordcolor},
    numberstyle=\tiny\color{gray},
    stringstyle=\color{red},
    basicstyle=\footnotesize\ttfamily,
    breakatwhitespace=false,         
    breaklines=true,                 
    captionpos=b,                    
    keepspaces=true,                 
    numbers=left,                    
    numbersep=5pt,                  
    showspaces=false,                
    showstringspaces=false,
    showtabs=false,                  
    tabsize=2,
    classoffset=0,
    morekeywords={void,char,int,uint64_t},
    keywordstyle=\color{keywordcolor}\bfseries,
    classoffset=1,
    morekeywords={if,else,for,return,struct,sizeof},
    keywordstyle=\color{keywordcolor2}\bfseries,
    classoffset=2,
    morekeywords={int,char},
    keywordstyle=\color{typecolor}\bfseries,
    classoffset=3,
    morekeywords={obj},
    keywordstyle=\color{functioncolor}\bfseries,,
    classoffset=0,
}
\newcommand{\percentagecolor}[1]{
}

\ifCLASSINFOpdf
\else
\fi
\begin{document}
%
% paper title
% Titles are generally capitalized except for words such as a, an, and, as,
% at, but, by, for, in, nor, of, on, or, the, to and up, which are usually
% not capitalized unless they are the first or last word of the title.
% Linebreaks \\ can be used within to get better formatting as desired.
% Do not put math or special symbols in the title.
\newcommand{\sys}{\textsc{ShadowBound}\xspace}

\newcommand{\todo}[1]{\textcolor{blue}{#1}}
\title{\Large \bf \sys: Efficient Heap Memory Protection Through Advanced Metadata Management and Customized Compiler Optimization \vspace{-10pt}}
% \author{Anonymous Author(s)}

% author names and affiliations
% use a multiple column layout for up to three different
% affiliations
\author{
{\rm Zheng Yu}\\
Northwestern University
\and
{\rm Ganxiang Yang}\\
Northwestern University
\and
{\rm Xinyu Xing}\\
Northwestern University
}
\maketitle

\subsection*{Abstract}

In software development, the prevalence of unsafe languages such as C and C++ introduces potential vulnerabilities, especially within the heap, a pivotal component for dynamic memory allocation. Despite its significance, heap management complexities have made heap corruption pervasive, posing severe threats to system security. While prior solutions aiming for temporal and spatial memory safety exhibit overheads deemed impractical, we present \sys, a unique heap memory protection design. At its core, \sys is an efficient out-of-bounds defense that can work with various use-after-free defenses (e.g. MarkUs, FFMalloc, PUMM) without compatibility constraints. We harness a shadow memory-based metadata management mechanism to store heap chunk boundaries and apply customized compiler optimizations tailored for boundary checking. We implemented \sys atop the LLVM framework and integrated three state-of-the-art use-after-free defenses. Our evaluations show that \sys provides robust heap protection with minimal time and memory overhead, suggesting its effectiveness and efficiency in safeguarding real-world programs against prevalent heap vulnerabilities.

\section{Introduction}\label{sec:intro}

%%% Say heap problem is very serious

In the domain of software development, a notable majority of programs are developed using unsafe languages, such as C and C++, which while granting detailed control over low-level operations and memory management, also invariably expose them to a range of vulnerabilities. The heap, acting as a critical component for dynamic memory allocation, assumes an essential role in these programs by managing data objects of sizes that are not fixed or predetermined. However, the complexity of heap management has made it a hotspot for vulnerabilities, with heap corruption emerging as a widespread issue \cite{sok_sanitizer, sok_mem}. Not only are heap vulnerabilities prevalent, but they also pose severe threats to system security, as they can be exploited to manipulate data, bypass security defenses, and execute arbitrary code, often having serious consequences for affected systems \cite{heelan2018automatic,wang2021maze,auto_heap_exp}. Consequently, comprehensive heap memory protection is imperative.

%%% Limitation of previous work

A comprehensive heap memory protection system should encompass both the temporal and spatial memory safety, ensuring that use-after-free and out-of-bound vulnerabilities are rendered completely unexploitable through its protection mechanisms. Numerous prior works \cite{asan,duck2018effectivesan,softbound,valgrind,pacmem,asan_junxu,zhang2021sanrazor} target this objective, all of which successfully provide both temporal and spatial memory safety. However, all of them suffer from an overhead exceeding 1.5x, rendering them impractical defenses in real-world programs. Additionally, some of them mainly used for bug detection or debugging, are not robust in terms of security \cite{sok_sanitizer}. Although these tools are not efficient, some prior works \cite{Ainsworth2020MarkUsDU,ffmalloc,Yagemann2023PUMMPU} can provide robust temporal memory safety with a low time overhead. Consequently, we believe it is possible to build a comprehensive memory protection system with them.

%%% intro of shadowbound and challenge

Inspired by these observations and thoughts, we propose \sys to provide a comprehensive heap memory protection mechanism. The core of \sys is an efficient out-of-bounds defense, designed to prevent the exploitation of out-of-bounds bugs by instrumenting boundary checks into programs. Moreover, it is capable of seamlessly integrating with various Use-After-Free (UAF) defenses without encountering compatibility issues. While the concept behind \sys is straightforward, developing such an out-of-bounds defense is not straightforward. We pinpoint two challenges that necessitate addressing. Firstly, the challenge lies in minimizing the time overhead of the out-of-bounds defense. Secondly, compatibility issues with UAF defenses present a hurdle. Most state-of-the-art UAF defenses necessitate the introduction of a new allocator. Thus, it is beneficial for the out-of-bounds defense to avoid any requirements for the allocation algorithm, consequently contracting the design space. Almost all previous works\cite{lowfat,TAILCHECK,kroes2018delta,sgxbound}, which focused on spatial memory safety as discussed in Section \ref{overview}, encountered similar challenges.

%%% How to address these challenges

We address these challenges through two perspectives. Initially, we designed a shadow memory-based metadata management mechanism to store the boundary of each pointer's corresponding heap chunk. This strategy ensures that \sys necessitates only a single load instruction and a handful of arithmetic instructions to extract boundaries during checks. Moreover, the shadow memory is orthogonal to the allocation algorithm, thereby enabling its integration with various allocators, specifically those utilized in UAF defense. Such a design lays the foundational framework for constructing an efficient and comprehensive heap memory protection mechanism.

Building upon the established foundation, we deployed a series of customized compiler optimizations, specifically tailored for boundary checking. Unlike traditional methods, which instrument boundary checking at the point of pointer dereference, \sys instruments these checks at the site of pointer arithmetic, thereby facilitating easier optimization. Generally, a compiler can retrieve more information at the pointer arithmetic site. At the pointer dereference site, the pointer might be passed externally, which hinders the compiler from fetching any of the pointer's properties. Conversely, the compiler can consistently utilize the computing process information at the pointer arithmetic site. We designed five optimizations based on this information, which significantly reduce the time overhead of \sys.

We implemented \sys atop LLVM 15 and integrated it with three state-of-the-art UAF defense mechanisms: PUMM\cite{Yagemann2023PUMMPU}, FFMalloc\cite{ffmalloc}, and MarkUs\cite{Ainsworth2020MarkUsDU}. To understand its security and practicality implications, \sys was applied to common benchmarks and real-world programs. Specifically, we utilized \sys to safeguard 19 programs against 34 exploitable out-of-bound bugs. With the protection provided by \sys, all exploits were successfully mitigated. Additionally, we undertook synthesis vulnerability testing, generating 244 inputs to trigger out-of-bounds bugs in various ways. Furthermore, \sys successfully prevented the triggering of these bugs. For performance evaluation, we assessed \sys using the SPEC CPU2017 and SPEC CPU2006 benchmark suites, as well as three real-world applications: Nginx, Chakra, and Chromium. 
% \zip{what does unmodified mean?} Unmodified 
\sys exhibits a 5.72\% and 10.58\% time overhead on SPEC CPU2017 and SPEC CPU2006, respectively, and introduces negligible overhead to the tested real-world programs.

%%% contributaion
In summary, this paper provides the following contributions:

\begin{itemize}

\item We introduce \sys, which employs a novel metadata design utilizing a compact method to encode boundary information in shadow memory. This approach enables \sys to quickly fetch the boundaries of a pointer, thereby making it compatible with various Use-After-Free (UAF) defenses and providing both spatial and temporal safety with minimal overhead.

\item We propose a series of novel optimization techniques customized for boundary checking at pointer arithmetic sites, significantly reducing time overhead. These optimizations have been implemented in \sys, based on LLVM 15. An ablation study demonstrates their effectiveness in reducing time overhead.

\item Through a thorough evaluation, \sys has been proven to offer robust spatial memory protection in various environments. It consistently maintains minimal time and memory overhead across a spectrum of benchmarks and real-world applications, affirming its full compatibility with three state-of-the-art UAF defenses.

\end{itemize}

\section{Background \& Design Overview}\label{overview}
% \zip{why it is problem definition not background}

\begin{table}[t]\small
    \tabcolsep=2.2pt
    \centering
    \begin{tabular}{cllc}
       \toprule
        & \textbf{Associated Address} & \textbf{Access} & \textbf{Exp?} \\
        \midrule
        \textbf{S1} & Inaccessible & Crash & \ding{54} \\
        \textbf{S2} & Accessible \& No Overlap & Benign & \ding{54} \\
        \textbf{S3} & Accessible \& Overlap & Data Leakage or Corruption & \ding{52} \\
        \bottomrule
    \end{tabular}
    \caption{Consequences of heap out-of-bounds bugs.}
    \label{tab:conseq}
\end{table}

\subsection{Heap Memory Protection}

A comprehensive heap memory protection mechanism should ensure both temporal and spatial memory safety for a program. In the realm of temporal memory safety, several Use-After-Free (UAF) defenses stand out for their remarkable performance, notably state-of-the-art solutions such as MarkUs\cite{Ainsworth2020MarkUsDU}, FFMalloc\cite{ffmalloc}, and PUMM\cite{Yagemann2023PUMMPU}. These tools provide full temporal memory safety, setting them apart from partial or probabilistic memory protection methods\cite{Archipelago, DieHarder, FreeGuard, Guarder}. Moreover, they outperform many of their predecessors. For instance, MarkUs leverages a garbage collection algorithm to identify live pointers, thereby preventing the creation of dangling pointers and offering improved performance compared to earlier solutions\cite{10.1145/3243734.3243826,Shin2019CRCountPI}. FFMalloc introduces an efficient one-time-allocation (OTA) allocator strategy, surpassing older solutions like Oscar\cite{oscar} and other OTA allocators\cite{1633516}. PUMM, in contrast, utilizes static analysis to pinpoint code units in charge of specific tasks, deferring the reallocation of memory freed by the active unit until it finishes its operations.

However, even with these advanced temporal memory protection mechanisms in place, the challenge of developing a comprehensive heap memory protection mechanism persists due to the absence of an adequate spatial memory protection mechanism. For instance, tools like MEMCHECK\cite{valgrind}, TAILCHECK\cite{TAILCHECK}, ASAN\cite{asan} and its variants\cite{asan_junxu, zhang2021sanrazor} are redzone-based spatial memory error detection systems, which are vulnerable to bypasses\cite{sok_sanitizer}. While ESAN\cite{lowfat, duck2018effectivesan}, SoftBound\cite{softbound}, and PACMEM\cite{pacmem} offer robust spatial memory protection, they still suffer from significant time overheads. Moreover, the designs of TAILCHECK, ASAN, and ESAN require a custom allocation algorithm to adjust the heap layout for out-of-bound checking. This can lead to conflicts when UAF defense introduces new allocators, such as MarkUs and FFMalloc. Although tools like DeltaPointer\cite{kroes2018delta} and SGXBound\cite{sgxbound} are relatively more efficient in terms of time overhead, they limit memory allocation to 4GB. Considering that UAF defense also amplifies memory overheads, this hinders their utility in large-scale applications.

\subsection{Out-Of-Bounds Exploitation}

Out-of-bounds vulnerability is a specific manifestation of spatial errors. Depending on the behavior of a program and the logic of the memory allocator, the consequences of a heap out-of-bounds bug can vary significantly. We present a comprehensive summary of these potential consequences in Table \ref{tab:conseq}, where the associated address refers to the memory address involved in an out-of-bounds read or write instruction. If the system has removed permission to access the associated address (S1), the out-of-bounds access will trigger a page fault and result in a crash. Another consequence is when the address remains accessible but still falls within the original heap chunk, without overlapping with other heap chunks or the freed regions (S2). This situation can occur because most allocators align the size of each heap chunk to 8 or 16 bytes, so they allocate more memory space than the program initially requested. Consequently, out-of-bounds access within such a region may happen. In the final scenario, the memory remains accessible, and the associated address may overlap with another region (S3), including other heap chunks, freed regions, or even extending beyond the heap boundary.

The bug in S1 is inherently non-exploitable, as it consistently results in a crash. In S2, despite the presence of an out-of-bounds bug within the logic of the program, such out-of-bounds access will never occur due to the extra space allocated by the allocator. In these instances, we can assume that the allocator "fixes" the bug. Conversely, an out-of-bounds error in S3 is notably susceptible to exploitation. An attacker can strategically manipulate the layout of the heap, causing the associated address to overlap with critical data. This could lead to data leakage, corruption, or even enable privilege escalation, compromising the entire system. To exploit an out-of-bounds bug in S3, the attacker must first create an out-of-bounds pointer and subsequently attempt to trigger a dereference with this pointer. If we can detect or eliminate the creation of out-of-bounds pointers that may be dereferenced later, any exploitation attempt will be rendered ineffective. \sys follows such a line of thought, which can defend against out-of-bounds exploitation by either transferring the S3 to S1/S2 or detecting all S3.

\subsection{Design Overview}\label{approach_overview}

\sys diligently works to prevent the exploitation of out-of-bounds bugs by instrumenting boundary checks into programs. It encompasses two principal modules: the metadata management module and the compiler module. The metadata management module maintains shadow memory, recording the boundaries of each pointer's corresponding heap chunk within the shadow memory at allocation sites. Concurrently, the compiler module instruments boundary-checking instructions at pointer arithmetic sites, precluding the generation of out-of-bounds pointers. Additionally, optimization is an essential role of the compiler module; it employs a suite of customized optimization techniques, specifically tailored for boundary checking, which ensures \sys incurs very low time overhead. Notably, both modules are meticulously designed to guarantee seamless integration with various UAF defenses, enabling \sys to collaborate with other UAF defenses to provide exhaustive heap protection.

\section{Threat Model}

\sys can be deployed either in conjunction with other UAF defense mechanisms or independently. When deployed alongside other UAF defenses, it is presumed that the target program contains one or more heap out-of-bounds and use-after-free vulnerabilities. If \sys is used independently, the assumption is limited to the presence of heap out-of-bounds vulnerabilities. In this threat model, an attacker can only attempt to exploit these vulnerabilities to potentially escalate privileges. Our goal is to prevent these vulnerabilities to being exploitable.

We assume the shadow memory will not compromise the security for three reasons. First, doing so would require exploiting vulnerabilities in our concise and robust shadow memory management code, which comprises less than 100 lines, giving us confidence in its security. Second, compromising the system in another way would necessitate exploiting a program vulnerability that allows an underflow write, overcoming a substantial 4TB space between the heap and shadow memory. This level of intrusion would require sophisticated exploitation primitives, which \sys is designed to detect and thwart preemptively. Finally, the attacker has no chance to infer the shadow memory addresses of objects. Since the attacker can only exploit vulnerabilities to leak some information, similar to the second reason then exploitation will be detected by \sys.

\section{Metadata Management}\label{sec:metadata}

In this section, we introduce our metadata management design, covering its layout, creation process, and how \sys leverages this metadata to enhance program security. We will also discuss the compatibility of metadata management with various UAF defense mechanisms.

\subsection{Shadow Memory Layout}\label{layout}

\begin{figure}[t]

\includegraphics[width=\columnwidth]{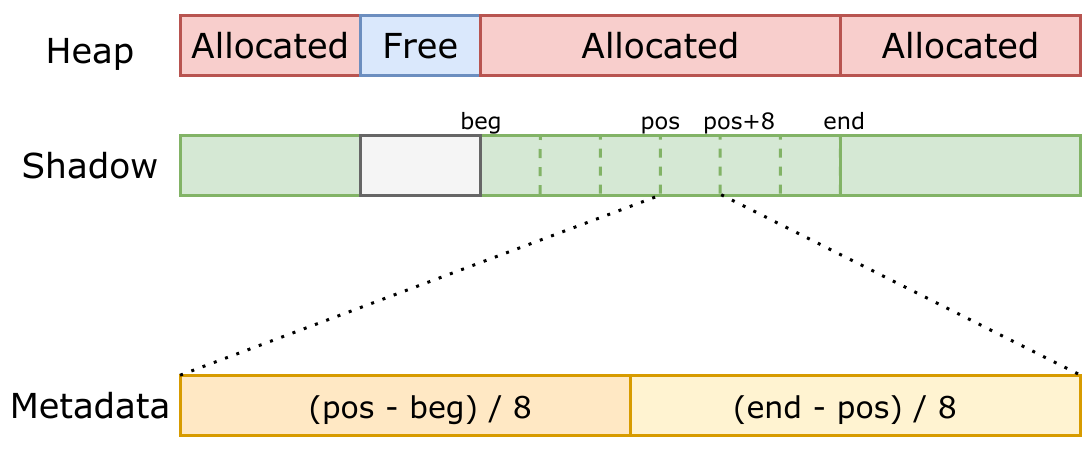}
\caption{Shadow Memory Mapping and Metadata Layout}
\label{fig:shadow}
\vspace{-10pt}
\end{figure}

\sys utilizes shadow memory to track the boundaries of each pointer's corresponding heap chunk in the program. As shown in Figure \ref{fig:shadow}, \sys maps each aligned 8 bytes of the program's heap memory into 8 bytes of shadow memory. Within this shadow memory representation, the front 4 bytes are dedicated to storing one-eighth of the length from the aligned 8 bytes to the beginning of the respective heap chunk. Conversely, the back 4 bytes are used to record one-eighth of the length from the aligned 8 bytes to the end of the corresponding heap chunk. The creation and initialization of shadow memory occur immediately following the allocation operation. The feasibility of this design stems from two fundamental observations.

Firstly, nearly all mainstream allocators, such as ptmalloc\cite{ptmalloc}, tcmalloc\cite{50370}, and jemalloc\cite{jemalloc}, default to 8-byte or 16-byte aligned allocations. This intrinsic behavior ensures that any valid 8-byte aligned memory address always belongs to the same memory chunk. Consequently, there is no need to maintain boundary information for every byte; instead, we can efficiently store the boundary information for every 8-byte aligned block. This approach significantly optimizes memory usage. The second observation is that the maximum single-time allocation size for these allocators is limited to 8 GB ($2^{33}$ bits). AddressSanitizer \cite{asan}'s default allocator also has such a feature. Building upon the previous observation, we can deduce that the length between any 8-byte aligned block and the beginning or end of the corresponding chunk is always a multiple of 8. Hence, 30 bits are sufficient to store this length information, and we can use 8 bytes or 64 bits to store both length values.

\begin{figure}[t]
\begin{lstlisting}[style=mystyle]
void foo(void *ptr, int n) {
    bound_check(ptr, ptr + sizeof(int));
    int *arr = (int *) ptr;

    for (int i = 0; i < n; ++i) {
        bound_check(arr, arr + i + 1);
        other_function(&arr[i]);
    }
}
\end{lstlisting}
\caption{Example: Instrumentation of \sys}
\label{lst:instrument}
\vspace{-10pt}
\end{figure}

\subsection{Boundary Checking}\label{boundary_checking}

As mentioned in subsection \ref{approach_overview}, unlike traditional checking methods that instrument checking instructions at the pointer dereference site, \sys instruments these checks at the pointer arithmetic site. LLVM provides two types of pointer arithmetic instructions. The \texttt{getelementptr} (GEP) instruction takes a base pointer and a set of indices to calculate the result pointer. If these indices are not properly checked, an attacker can exploit a GEP to perform an out-of-bounds access. The \texttt{bitcast} instruction (BC) allows the conversion of a base pointer from one type to a result pointer with another type. However, if the pointer cannot sufficiently accommodate the target type, it may potentially result in an overflow issue. Listing \ref{lst:instrument} provides an example of how \sys checks these two types of instructions. In line 3 of the code, an attempt is made to convert a \texttt{void*} pointer to an \texttt{int*} pointer. To ensure that the resulting pointer can hold at least one integer, \sys inserts a check at line 2. In line 6, a new pointer is generated, and \sys inserts a check to ensure that the new pointer and the old pointer are located within the same memory chunk.

The pseudocode for the boundary check, as shown in Listing \ref{lst:bound_check}, describes a checking function that takes two parameters. The function first checks if the pointer is located in the heap region (Line 2). If so, it aligns the base pointer to an 8-byte boundary and calculates the corresponding shadow memory address (Lines 3-4). It then loads the metadata from the shadow memory (Line 5). The next two lines calculate the boundary. Specifically, for the beginning address, the lower 32 bits of the metadata are taken and shifted left by 3 (effectively multiplying by 8) to get the actual distance from the aligned address to the beginning of the corresponding heap, and then this distance is subtracted from the aligned address to determine the beginning address (Line 6). Conversely, for the end address, the upper 32 bits are taken and shifted left by 3 to get the distance from the aligned address to the end of the corresponding heap, then this distance is added to the aligned address to obtain the end address (Line 7). Finally, the function verifies whether the result pointer falls outside these computed boundaries (Line 8). If the result pointer is less than the starting address or greater than or equal to the ending address, it triggers an error, leading to program termination. Due to the boundary-checking code's reliance on just a single load instruction and a handful of arithmetic operations, the verification process is exceptionally quick, resulting in minimal execution time.

\begin{figure}[t]
\begin{lstlisting}[style=mystyle]
void bound_check(uint64_t old, uint64_t res) {
    if (!IsHeapAddress(old)) return;
    uint64_t align = old & ~7;
    uint64_t shadow = align + OFFSET;
    uint64_t pack = *(uint64_t*) shadow;
    uint64_t beg = align - ((pack & 0xffffffff) << 3);
    uint64_t end = align + ((pack >> 32) << 3);
    if (res < beg || res >= end)
        error("Heap out-of-bounds Detected");
}
\end{lstlisting}
\caption{Pseudocode for boundary checking}
\label{lst:bound_check}
\vspace{-10pt}
\end{figure}

\subsection{Compatibility with UAF Defense}\label{comp_uaf}
\sys introduces continuous memory as shadow memory to track the boundaries of each heap chunk. This technique empowers \sys to seamlessly integrate with virtually all UAF defense. Within the scope of this research paper, we have successfully integrated three state-of-the-art UAF defenses into \sys, MarkUs\cite{Ainsworth2020MarkUsDU}, FFMalloc\cite{ffmalloc}, and PUMM\cite{Yagemann2023PUMMPU}. Each of these defenses implements a secure allocator, which means they need to modify the allocation algorithms, potentially resulting in changes to the heap layout. It's noteworthy that the shadow memory operates independently of the allocation algorithm and heap layout. This ensures that \sys does not conflict with the design of secure allocators.

In addition to secure allocators, some research employs pointer invalidation algorithms for UAF defense \cite{Lee2015PreventingUW,dangsan,Younan2015FreeSentryPA}. While this approach may not currently provide optimal performance, there is potential for improvement in its effectiveness in the future. Consequently, we also demonstrate \sys's compatibility with these tools. Typically, these approaches involve tracking all pointers to allocated objects and explicitly invalidating them once the referenced objects are freed, requiring program instrumentation. It's worth noting that \sys also necessitates program instrumentation, but there is no conflict because \sys instruments pointer arithmetic instructions, while they only need to instrument dereference instructions.

\section{Compiler Optimization}\label{sec:opt}

In this section, we introduce the compiler optimization employed by \sys, which play a important role in reducing time overhead while maintaining robust security.

\begin{figure}
\begin{lstlisting}[style=mystyle]
struct obj {
    int x, y, z;
};
void foo() {
    char *c = malloc(3 * sizeof(int));
    struct obj* o = malloc(sizeof(struct obj));
    ...
}
void bar(char *c) {
    c[0] = 'x';
    c[1] = 'y';
    c[2] = 'z';
    escape(c + 1);
}
void zoo(struct obj *o) {
    o->x = 1;
    o->y = 2;
    o->z = 3;
}
\end{lstlisting}
\caption{Example of Runtime-Driven Checking Elimination}
\vspace{-10pt}
\label{lst:rt-driven}
\end{figure}

\subsection{Runtime-Driven Checking Elimination}

This optimization is based on a simple idea: if each heap chunk has infinite space, out-of-bounds access becomes impossible, rendering all boundary checks redundant and eliminable. The core of this concept relies on the runtime environment's ability to provide the compiler with extra information, enabling it to eliminate specific boundary checks that static analysis techniques alone cannot resolve. By leveraging runtime information, certain optimizations become feasible, empowering the compiler to remove unnecessary boundary checks, and significantly enhancing system efficiency. However, it's impractical to allocate infinite or even very large spaces for every chunk due to the potential for high memory overhead. Therefore, \sys chooses an improved approach to balance time overhead and memory overhead. Specifically, \sys reserves a fixed $n$ bytes for every heap chunk, denoted as reserved space. Then, \sys will try to find all eliminable boundary checks using the reserved space provided by the runtime.

As mentioned in subsection \ref{boundary_checking}, \sys inserts boundary checks for every pointer arithmetic instruction. Each instruction contains a base pointer argument and generates a result pointer. These two pointers are used as the arguments passed to the boundary checking. Notably, if the offset between the result pointer and base pointer can be confirmed to be less than $n$ bytes at compile time, and the result pointer will never be used as a base pointer in another boundary checking, \sys can safely remove the boundary checking. This is because \sys already reserves $n$ bytes for every heap chunk, ensuring that every live pointer, which may be used as a base pointer in boundary checking, has at least $n$ bytes of space available.

We can use Listing \ref{lst:rt-driven} as an example to illustrate how the optimization works. In the function \texttt{bar}, the argument \texttt{c} generates three pointers in lines 10 to 12, all with offsets less than 8. If we set $n$ to 8, \sys can safely remove these checks. However, we cannot eliminate the check for line 13 because the pointer \texttt{c + 1} is passed to another function, indicating that it may potentially be used as a base pointer for boundary checking in that function. If \sys were to eliminate the boundary check for a pointer that has already pointed to the reserved space, it could lead to false negatives due to the elimination in the \texttt{escape} function.

Furthermore, the optimization concept can also be applied to eliminate structure member accesses, as demonstrated in the function \texttt{zoo} of Listing \ref{lst:rt-driven}. The object is allocated at line 6, and the return type of the allocation function is \texttt{void*}, not \texttt{struct obj*}. Consequently, the compiler inserts a type-casting instruction, which prompts \sys to insert a boundary check to ensure that the allocated memory space can accommodate the structure \texttt{obj}. This type-casting validation happens in runtime and allows the compiler to confidently determine that the memory space associated with a typed pointer is at least the size of its type. Consequently, the compiler can infer that pointers referring to structure fields are in-bounds. As a result, \sys can safely eliminate the boundary checks for lines 16 to 18.

\begin{figure}
\begin{lstlisting}[style=mystyle]
struct obj {
    int *a;
    int len;
};
void foo(struct obj *o, int len) {
    // initialize the object
    o->len = len;
    o->a = malloc(len * sizeof(int));
    ...
void bar(struct obj *o) {
    for (int i = 0; i < o->len; ++i)
        o->a[i] = i;
}
\end{lstlisting}
\caption{Example of Security Pattern Identification and Directional Boundary Checking Optimization}
\label{lst:pattern}
\end{figure}

\subsection{Directional Boundary Checking}

The boundary checking of \sys consists of two parts: the underflow check ensures that the memory access address is greater than or equal to the lower boundary of the allocated memory chunk, while the overflow check verifies that the address falls within the upper boundary. We can observe that during boundary checking, the base pointer is always valid. If not, \sys will throw an error when creating the base pointer and the program terminated. So that if we can determine the sign (positive or negative) of the offset between the base pointer and the result pointer, we can optimize by inserting only one side of the checking code. For instance, in line 12 of Listing \ref{lst:pattern}, the offset variable \texttt{i} starts from zero, ensuring that it is always positive, eliminating the need to check for potential underflow pointers.

To implement this optimization, \sys leverages LLVM's range analyzer \cite{llvmse}, a powerful tool that provides insights into the possible value ranges of variables during program execution. By analyzing the sign of the offset, we can determine which boundary checks cannot be violated. For instance, if range analysis indicates a consistently positive offset, we optimize by focusing solely on the overflow check, thus eliminating the need for underflow verification. It is worth emphasizing that the use of range analysis also ensures that checked arithmetic instructions will not result in integer overflow (or wraparound), thereby maintaining the security assumption of \sys. This approach is particularly beneficial in sections of code that are loop-intensive, as loops tend to undergo multiple iterations with consistent direction.

\subsection{Security Pattern Identification}
Many programs frequently employ simple code patterns to prevent out-of-bounds, making it unnecessary to validate code that has already been safeguarded by the original program. In this context, \sys focuses on the identification of two primary patterns: Constant Array Argument and Length-Pointer Association.

\paragraph{Constant Array Argument}

The pattern is used to describe function arguments, where specific function arguments consistently accept constant-length arrays and the access offset for these arguments is always bound by the same constant. These arguments can be either stack arrays, global arrays, heap arrays, or constant-length arrays within a structure. If the arguments always belong to the first two types, we can directly remove the associated checks, as \sys primarily focuses on heap security. To accomplish this, \sys employs whole-program analysis to identify these function arguments and subsequently eliminates the redundant boundary checks associated with them.

The identification algorithm focuses on functions that are not the target of indirect call instructions. This is because if a function can be the target of an indirect call, it becomes challenging to analyze which properties the function's arguments consistently hold, as we may not be able to identify all call sites for this function. To determine if a function is the target of indirect call instructions, \sys examines every instruction in the program and verifies that the function only appears as the called-function argument in call-related instructions. If the function is stored in memory or passed as a function argument, \sys assumes it might become the target of an indirect call.

After identifying eligible functions, \sys inspects the pointer-type arguments of each function. It traverses the entire program to identify all call sites associated with each function. \sys evaluates each call site to determine whether the function's arguments consistently exhibit the characteristics of a constant length. To simplify the analysis, this assessment revolves around two key criteria: firstly, whether the pointer is instantiated within the same function where the call site is located, and secondly, whether the length of the pointer remains constant. If an argument is discovered to be loaded from memory or supplied as a function argument, \sys refrains from engaging in further recursive analysis and instead assumes that it fails to meet the requisite criteria.

If an argument consistently meets the criteria of being a constant-length pointer after assessing all call sites, \sys proceeds to the final step. In this concluding phase, \sys meticulously examines the pointer arithmetic instructions related to the argument within the corresponding function. It leverages the LLVM range analyzer to confirm that the offsets generated by these instructions are strictly bound by the corresponding constant length. For those arguments that successfully pass all verifications, \sys can confidently eliminate the associated checks.

\paragraph{Length-Pointer Association} 

The Length-Pointer Association pattern is a common structural configuration in programs, characterized by two essential members: one representing a pointer and the other indicating the length of the data pointed to. These two members are referred to as the pointer member and the length member in the pattern. This pattern establishes a direct relationship between the pointer member and the corresponding length member. As demonstrated in Listing \ref{lst:pattern}, through the initialization code at lines 7 to 8 for the structure \texttt{obj}, we can discern that the structure \texttt{obj} exemplifies this pattern, with \texttt{a} serving as a pointer to an integer array and \texttt{len} denoting the length of that array. 

Identifying the Length-Pointer Association pattern is not always straightforward. For example, the modification for the pointer member and length member may not happen in the same scope, which may require heavy dataflow analysis techniques to address it. To simplify the identification process, we have introduced specific constraints that facilitate recognition. Firstly, both members must consistently undergo modifications within the same scope. Secondly, modifications to these two members should exclusively occur at the allocation or deallocation site. This typically means that the pointer is assigned as the return value of an allocation function or as a null pointer. Finally, the length field must precisely match the allocated memory's size at the allocation site. These constraints streamline the pattern identification process.

\sys follows a systematic process to identify the Length-Pointer Association pattern. Initially, it identifies all allocation function call sites and checks if their return values are assigned to structure members. If this condition is met, \sys further verifies if the length argument of the allocation function call is stored in another structure member. If both conditions are satisfied, \sys assumes the potential presence of the pattern. Subsequently, \sys scans the entire program to determine whether the two members of the structure fulfill all the pattern requirements that we mentioned before at all modification sites.

Upon confirming that these two members match the pattern, it means that the length member consistently indicates the length of the pointer member, \sys proceeds to check all pointer arithmetic instructions that utilize the pointer member as the base pointer. If the offset between the base pointer and the result pointer is safeguarded by the length member, \sys confidently removes the corresponding boundary checks. This meticulous process ensures precise optimization based on the Length-Pointer Association pattern.

\begin{figure}
\begin{lstlisting}[style=mystyle]
void foo(char *p) {
    char *a = p + 1;
    char *b = a + 2;
    char *c, *d;
    if (random() > 0.5)
        c = a + 3;
    else
        c = b + 4;
        
    for (d = c; d < p + 100; d++)
        *d = 'x';
}
\end{lstlisting}
\caption{Example of Merge Metadata Extraction}
\label{lst:merge}
\end{figure}

\subsection{Merge Metadata Extraction}

Within its boundary-checking process, \sys employs a two-stage approach, which involves an initial \textit{extraction} stage to obtain the base pointer's boundary, followed by a subsequent \textit{checking} stage to validate the result pointer. It's essential to highlight that the extraction stage is notably more time-consuming than the second stage due to the requirement of a load instruction for extracting metadata from the shadow memory. Importantly, the load address is exclusively determined by the base pointer. Consequently, in cases where multiple result pointers are computed from the same base pointer, it becomes possible to optimize by merging the extraction stage of these result pointers' boundary-checking processes, thereby avoiding redundant address loading.

To maximize the merging of extraction stages, \sys employs a backtrace algorithm. Specifically, \sys enumerates all pointer arithmetic instructions and attempts to backtrack their sources. We illustrate how the algorithm works using Listing \ref{lst:merge}. We start with the pointer \texttt{d}, which is initialized with \texttt{c} and then self-incremented, making \texttt{d} dependent on \texttt{c}. While \texttt{c} could be computed from either \texttt{a} or \texttt{b}, it is evident that both \texttt{a} and \texttt{b} are ultimately derived from \texttt{p}. Consequently, all pointers in the function \texttt{foo} are generated from \texttt{p}, establishing \texttt{p} as the source of pointers \texttt{a, b, c}, and \texttt{d}. Once all pointer sources are collected, for those pointers sharing the same source, we insert the boundary extraction code after the source. This eliminates the need to extract the boundary again in the subsequent checking.

% \begin{algorithm}[t]
% \small
% \textbf{Input:} A pointer $P$ \;
% \textbf{Output:} A set, denoted as $S$, which contains pointers computed from $P$ \;
% \textbf{Initialize:} WorkList = list(); Visited = set() \;
% WorkList.push($P$) \;
% \While{$WorkList.size() > 0$}{
%     $V$ = WorkList.pop() \;
%     \If {$V\in$ Visitied} {
%         \textbf{continue};
%     }
%     Visited.add($V$) \;
%     \ForEach{$I$ $\in$ All instruction that use $V$}{
%         \If {$I$ is GEP or BitCast instruction} {
%             WorkList.push($I$) \;
%             $S$.add($I$);
%         } 
        
%     }
% }
% \caption{Base Pointer Propagation}
% \label{algo:ptr_prop}
% \end{algorithm}

\subsection{Redundant Checking Elimination}

In this subsection, we introduce two common techniques employed by \sys to eliminate redundant checks. The first optimization typically occurs when both the allocation and the pointer arithmetic take place within the same scope, or when the base pointer is an array member of a structure. In such cases, the size of the base pointer is inherently determined by its allocation size or type. Consequently, the compiler can optimize this scenario by comparing the allocation size with the offset between the base pointer and the result pointer. If the offset is consistently smaller than the allocation size, \sys can confidently assert that a pointer arithmetic will never result in an out-of-bounds condition. This empowers \sys to eliminate the corresponding checks effectively.

The second technique entails conducting a thorough analysis of the relationships between pointer arithmetic instructions within a program. In particular, it commences by identifying cases where one pointer arithmetic instruction either dominates or post-dominates another instruction, provided that they share the same base pointer. In such instances, \sys will proceed to compare the offsets of these two pointer arithmetic instructions. If it is consistently observed that the dominated computing instruction's offset is greater than that of the other instruction, then the underflow checking for this instruction can be safely eliminated. Conversely, if it is consistently found that the dominated computing instruction's offset is smaller than that of the other instruction, then the overflow checking for this instruction can be safely eliminated.

\section{Implementation}

In this section, we describe our implementation of the compiler and runtime support for \sys, as well as our integration of state-of-the-art UAF defense mechanisms into the \sys framework.

\subsection{Compiler \& Runtime Support}

\sys is build upon LLVM 15.0.6 framework \cite{1281665} and consists of two components. The first, known as the function pass, is responsible for the insertion and optimization of checking instructions at the LLVM Intermediate Representation (IR) level. The second component is the runtime module, designed for shadow memory allocation and the management of associated metadata. These two components work seamlessly together, forming the foundational architecture of \sys.

\paragraph{Function Pass.}  The function pass is implemented as an internal sanitizer within LLVM. Unlike other sanitizers, such as AddressSanitizer \cite{asan} or Memory Sanitizer \cite{7054186}, this tool is not designed to detect a specific class of bugs. Instead, it serves as a defense mechanism. Users can enable \sys using the \texttt{-fsanitize=overflow-defense} flag. The function pass collects all pointer arithmetic instructions, including \texttt{getelementptr} and \texttt{bitcast}, and employs optimization algorithms that are mentioned in section \ref{sec:opt} to determine which instructions require checking and how they should be checked.

Given that various optimization methods may interact with each other, an improper optimization sequence could potentially hinder both optimization performance and compilation speed. Therefore, we meticulously select the order in which each optimization method is applied, ensuring that our choices result in the best possible outcome. The function pass begins with Runtime-Driven Checking Elimination, as it effectively eliminates a significant number of checking instructions, reducing the burden for subsequent optimizations. Following that, the function pass executes Redundant Checking Elimination, Security Pattern Identification, and Directional Boundary Checking Optimization in sequence. Finally, the function pass concludes with the Merge Metadata Operation.

It is noteworthy that Security Pattern Identification necessitates a whole-program analysis. However, performing a whole program analysis requires access to information from all functions within the program, whereas the LLVM function pass is limited in its scope, and capable of retrieving information only from the currently processed function. To address this challenge, we have implemented an external analyzer similar to KINT \cite{180262}. This analyzer is capable of both dumping and analyzing the IR code of each function, subsequently saving the analysis results to a configuration file before compiling the program. The function pass then accesses and utilizes the information stored in this configuration file for its optimization, ensuring a more holistic approach to security pattern identification and mitigation.

\paragraph{Runtime Module.}

The runtime module of \sys has two primary functions. First, it manages metadata, which includes the allocation of shadow memory. The size of the shadow memory in \sys is equal to that of the original heap, mirroring the configuration of MemorySanitizer. Thus, our implementation closely resembles MemorySanitizer's shadow allocation method. Moreover, we introduce reserve operations at allocation sites, as detailed in Section \ref{sec:opt}. Second, the runtime module checks the arguments of frequently used libc functions, such as \texttt{memset} and \texttt{strcpy}, to guard against heap out-of-bounds incidents within these functions. To instrument the necessary checking code for these libc functions, we use a technique similar to AddressSanitizer. This method effectively identifies the functions needing validation and inspects their arguments for potential out-of-bounds access.

\subsection{UAF Defense Integration}

As discussed in subsection \ref{comp_uaf}, \sys theoretically can collaborate with three state-of-the-art UAF defense tools: MarkUs, FFMalloc, and PUMM. In this subsection, we delve into the implementation of each tool and present how we seamlessly integrate them within the \sys framework, showcasing the integration achieved with minimal effort.

Both MarkUs and FFMalloc introduce new allocators. However, we encountered a challenge in which the heap region utilized by these allocators conflicted with our shadow memory region. Specifically, we designate two separate regions for the heap and shadow memory. The original heap memory of FFMalloc and Markus is not allocated within these regions. Therefore, we modify the code to adjust the heap region. It is important to note that we do not alter their allocation algorithm; we merely "shift" the allocated objects. Furthermore, we implemented code to facilitate the allocation and initialization of shadow memory after each allocation function. This adjustment ensured that these allocators gained the ability to allocate shadow memory for each heap chunk and initialize it.

The implementation of PUMM differs from that of MarkUs and FFMalloc. Instead of creating a new allocator, PUMM wraps the original allocator. This wrapping mechanism involves deferring specific deallocation operations based on a policy. PUMM generates this policy through binary-based analysis, which obviates the necessity for many modifications to enable integration. To integrate PUMM, our approach entails compiling the program with \sys and then utilizing PUMM to analyze the binary and generate the policy. Subsequently, PUMM can wrap the program's allocator by the generated policy.

\begin{table}[t]\small
    \tabcolsep=4pt
    \centering
        \begin{tabular}{llll}
         \toprule
           \normalsize{\textbf{CVE/Issue ID}} &
            \normalsize{\textbf{Link}} & 
            \normalsize{\textbf{Program}} & 
            \normalsize{\textbf{Prevention Type}}
            \\
         \midrule
            CVE-2021-32281   & \cite{CVE-2021-32281}  & gravity       & \ding{52} OOB Detected    \\
            CVE-2021-26259   & \cite{CVE-2021-26259} & htmldoc       & \ding{52} OOB Detected    \\
            CVE-2020-21595   & \cite{CVE-2020-21595} & libde265      & \ding{52} OOB Detected    \\
            CVE-2020-21598   & \cite{CVE-2020-21598} & libde265      & \ding{52} OOB Detected    \\
            CVE-2018-20330   & \cite{CVE-2018-20330} & libjpeg-turbo & \ding{52} OOB Detected    \\
            CVE-2021-4214    & \cite{CVE-2021-4214} & libpng        & \ding{52} OOB Detected    \\
            CVE-2020-19131   & \cite{CVE-2020-19131} & libtiff       & \ding{52} OOB Detected    \\
            CVE-2020-19144   & \cite{CVE-2020-19144} & libtiff       & \ding{52} OOB Detected    \\
            CVE-2022-0891    & \cite{CVE-2022-0891} & libtiff       & \ding{52} OOB Detected    \\
            CVE-2022-0924    & \cite{CVE-2022-0924} & libtiff       & \ding{52} OOB Detected    \\
            CVE-2020-15888   & \cite{CVE-2020-15888} & Lua           & \ding{52} OOB Detected    \\
            CVE-2022-0080    & \cite{CVE-2022-0080} & mruby         & \ding{52} Benign Running       \\
            Issue-5551       & \cite{mruby_issue_5551} & mruby         & \ding{52} Transformation \\
            CVE-2019-9021    & \cite{CVE-2019-9021} & php           & \ding{52} OOB Detected    \\
            CVE-2022-31627   & \cite{CVE-2022-31627} & php           & \ding{52} OOB Detected    \\
            CVE-2021-3156    & \cite{CVE-2021-3156} & sudo          & \ding{52} Benign Running       \\
            CVE-2022-28966   & \cite{CVE-2022-28966} & wasm3         & \ding{52} OOB Detected    \\     
        \bottomrule
        \end{tabular}
        \caption{Heap out-of-bounds Prevention Results for \sys on Real-World Vulnabilities.}
        \label{tab:realworld-applications}
\end{table}

\section{Evaluation}

In this section, we address three key questions to assess the effectiveness and efficiency of \sys:

\begin{enumerate}
  \item Does \sys effectively prevent heap out-of-bounds-based exploits?
  \item What is the quantitative impact of \sys on both runtime performance and memory utilization?
  \item How do individual customized compiler optimization algorithms influence time overhead?
\end{enumerate}

All the experiments were conducted on a bare-metal machine configured with Ubuntu 22.04 system, 12th Gen Intel i7-12700 CPU at 4.9 GHz, 32GB RAM, and 1T SSD storage.

\subsection{Security Evaluation}

\paragraph{Real-World OOB Prevention} To validate \sys's ability to prevent real-world heap out-of-bounds-based exploits, we conducted a comprehensive evaluation. This assessment included 34 real-world vulnerabilities; half of them were collected from previous works \cite{zhang2021sanrazor, asan_junxu} and \textsc{Magma}\cite{magma} dataset, while the others were selected from the CVE database \cite{cve_details,cve_program,vuln_db} and program-specific issue reports. These vulnerabilities spanned a diverse set of 19 applications, encompassing servers, video encoders, language interpreters, widely adopted libraries, and UNIX utilities. For each case, we collected the corresponding Proof of Concept (PoC) that triggers program crashes. We observed how these programs, when compiled with \sys, prevented the associated exploitation attempts.

In Table \ref{tab:realworld-applications}, we present the evaluation results on vulnerabilities selected by us. The results based on vulnerabilities collected from prior works are shown in Table \ref{tab:prior-realworld-applications} in the Appendix. \sys successfully prevents all cases, demonstrating its capability to prevent real-world heap out-of-bounds-based exploitation. In each case, we observed the three distinct forms through which \sys effectively prevented vulnerabilities: \textit{OOB Detected}, \textit{Benign Running}, and \textit{Transformation}:

\begin{itemize}
    \item \textbf{OOB Detected (OD)}: \sys successfully detects heap out-of-bounds read or write instructions when an attempt is made to compromise a program used as a proof-of-concept (PoC). In simpler terms, \sys acts as a vigilant guard, spotting and halting any unauthorized attempts to access memory beyond its boundaries.

    \item \textbf{Benign Running (BR)}: In certain scenarios, \sys "fixes" the vulnerability by allocating additional memory space for each memory chunk. This extra space transforms what would otherwise be an unauthorized memory access into a legitimate operation, similar to creating an additional safety buffer. We confirmed this using a semi-automatic gdb script. The script fetched both the original requested size and the actual allocated size to determine the address range of the reserved space, and then checked if the out-of-bounds access fell within this range.

    \item \textbf{Transformation (TF)}: \sys transforms heap out-of-bounds vulnerabilities into alternative, non-exploitable forms. During our manual verification process, we observed instances of null-pointer dereferences in these cases. These dereferences occurred because PoCs attempted to dereference pointers loaded from the reserved memory space created by \sys for each memory chunk, and the reserved space is cleared to zero during allocation. Thus, the null-pointer dereference happened.
    
\end{itemize}

\begin{table}[t]\small
    \centering
    \begin{tabular}{ccccc}
         \toprule
             \textbf{Program} & 
             \textbf{\#Input} & 
             \textbf{\#OD} & 
             \textbf{\#BR} & 
             \textbf{\#TF} 
             \\             
        \midrule
        cxxflit & 1 & 1 & 0 & 0 \\
        libpcap & 4 & 2 & 2 & 0 \\
        libxml2\_reader & 127 & 127 & 0 & 0 \\
        libxml2\_xml & 61 & 46 & 15 & 0 \\
        proj4 & 3 & 0 & 3 & 0 \\
        zstd & 48 & 45 & 3 & 0 \\
        \midrule
        \textbf{Total} & 244 & 221 & 23 & 0 \\
        \bottomrule
    \end{tabular}
    \caption{Heap out-of-bounds Prevention Results for \sys on Synthesis 
    Vulnerabilities.}
    \label{tab:vuln_injection}
\end{table}

\begin{figure*}[t]
    \centering
    \includegraphics[width=\textwidth]{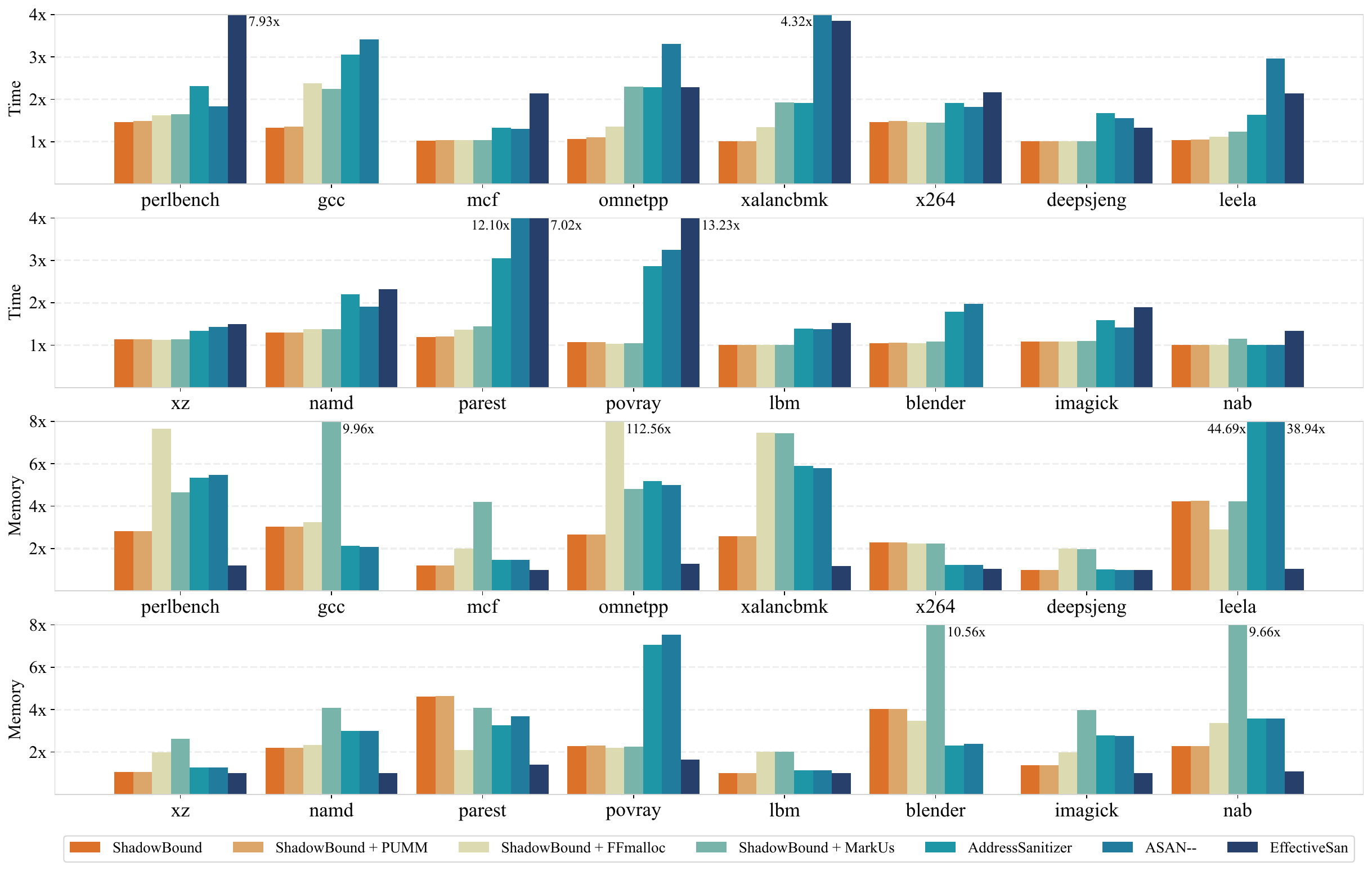}
    \caption{The time and memory overhead of \sys, \sys + PUMM, \sys + FFMalloc, \sys + MarkUs, AddressSanitizer, ASAN$--$, ESAN on SPEC CPU2017. 
    The upper two lines represent the results for time overhead, while the lower two outline the results for memory overhead. 
    A value of 1x signifies no overhead. 
    The geomean time overhead of each system is 5.72\%, 6.60\%, 9.95\%, 16.20\%, 62.03\%, 79.85\% and 138.76\%. The geomean memory overhead of each system is 
    54.59\%, 55.29\%, 218.20\%, 302.51\%, 116.55\%, 112.33\%, 2.70\%.}
    \label{fig:spec2017}
\end{figure*}

\paragraph{Synthesis Vulnerability Prevention} 

To demonstrate \sys's robustness in security, we conducted an extensive evaluation using the RevBugBench dataset \cite{FixReverter}. The dataset was originally designed for fuzzing testing and comprises a series of programs, each of which has been injected with over 1,000 distinct vulnerabilities collected from the real world. We employed AFL++\cite{AFLplusplus-Woot20} to fuzz each program for 24 hours, generating numerous inputs capable of triggering program crashes. Subsequently, we assessed the prevention ability by using these inputs. It's essential to clarify that although the vulnerabilities were sourced from real-world cases, they might not appear at the same time in the real world. Therefore, these vulnerabilities triggered by inputs are synthetic but they are very close to real-world vulnerabilities.

Furthermore, recognizing that not all inputs result in heap out-of-bounds errors and certain inputs might execute the same program path, we implemented input refinement using the following methodology. Initially, we employed ASAN to discern inputs that cannot trigger heap out-of-bounds errors. For instance, inputs that solely lead to stack out-of-bounds or null-pointer dereference issues were automatically filtered out. 
% Subsequently, we conducted a comparative analysis of the bug sets triggered by pairs of inputs. If two inputs produced identical bug sets, it indicated that they could induce a crash in almost the same manner. 
Notably, every input will be considered unique if it has a distinct execution path leading to the final crash site. This is because even if another input results in the same crash site, their execution paths may diverge.
Consequently, one of the redundant inputs was removed. The identification of triggered bug sets was facilitated by RevBugBench when provided with the program and inputs.

Following the meticulous cleanup process, we confirmed that all remaining inputs are indeed able to trigger heap out-of-bounds errors, with each input following a distinct trigger path. Subsequently, we compiled the benchmark programs with \sys and conducted testing using these inputs. The results, as detailed in Table \ref{tab:vuln_injection}, showcased \sys's effective prevention of these errors, with the majority being detected (\#OD), while the remaining inputs allowed the program to execute without issues (\#BR). All the benign running cases were also confirmed by the gdb script mentioned previously. Unlike our observations in real-world bug evaluations, we did not encounter any instances of \textit{Transformation} (\#TF), which we believe is normal given its dependency on the underlying program's logic.

\begin{figure}[t]
    \centering
    \includegraphics[width=\columnwidth]{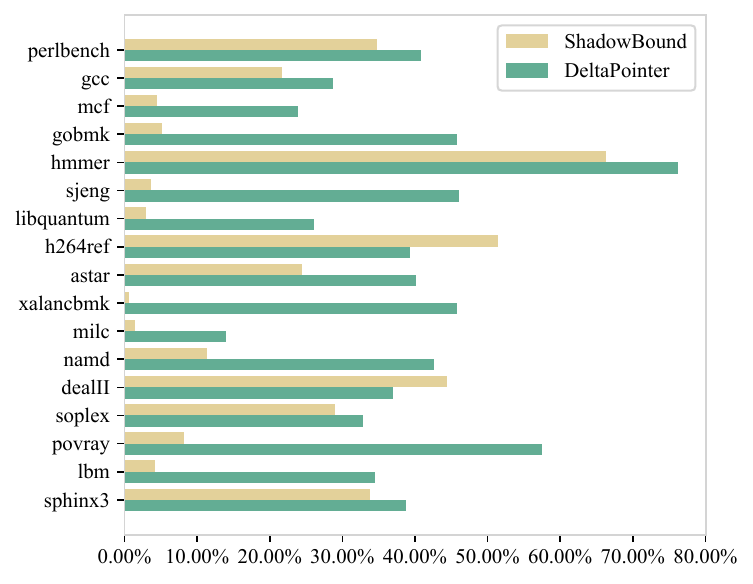}
    \caption{The runtime overhead of \sys and DeltaPointer on SPEC CPU2006, the geometric mean overhead of each system is 10.58\% and 37.08\%. }
    \label{fig:spec2006}
\end{figure}

\subsection{Performance Comparison}\label{perf-comp}

\paragraph{SPEC CPU Benchmark} 
To perform a comparative analysis between \sys and related tools that offer comprehensive heap memory protection, including both out-of-bounds (OOB) and Use-After-Free (UAF) defense, we integrated three state-of-the-art UAF defense tools: PUMM, FFMalloc, and MarkUs with \sys. We then conducted an extensive evaluation, including benchmarking against established tools such as ESAN, AddressSanitizer, MEMCHECK, and ASAN$--$, using the SPEC CPU2017 suite \cite{spec2017}. To provide clarity regarding the overhead introduced by \sys itself, we also present results for \sys performance on SPEC CPU2017. Furthermore, we compared \sys with DeltaPointer, the state-of-the-art tool primarily focused on out-of-bounds detection. However, it's worth noting that DeltaPointer exclusively supports SPEC CPU2006 \cite{spec2006}, and our attempts to migrate it to SPEC CPU2017 were unsuccessful. Consequently, our comparison between \sys and DeltaPointer is based on SPEC CPU2006. Regrettably, due to code compatibility issues, we had to exclude \texttt{471.omnetpp} and \texttt{401.bzip2} from the SPEC CPU2006 cases, as LLVM 15 failed to compile these benchmarks due to these codes are too old. To ensure fairness in our comparisons, we configured all tools to disregard detected errors, preventing premature termination. Additionally, we only enabled ASAN$--$, ESAN, and AddressSanitizer's heap error detector. In our evaluation, we excluded PACMem and SGXBound due to their reliance on specialized hardware for heap memory error detection, as well as the unavailability of their source code. For each tool, we executed the benchmarks ten times and reported the average results to mitigate the impact of randomness. 

Figure \ref{fig:spec2017} presents the results obtained from evaluating SPEC CPU2017 benchmarks. Notably, ESAN encountered issues and failed to execute the \texttt{gcc} and \texttt{blender} benchmarks. As for MEMCHECK, it exhibited a significant performance slowdown of over 4x in every test case, its geomean time and memory overhead is 2426.66\% and 92.95\%, rendering its results unfit for inclusion in the figure. Furthermore, it is worth highlighting that both ASAN$--$ and ESAN's results were inferior to AddressSanitizer. This discrepancy is attributed to the fact that AddressSanitizer in LLVM 15 deployed more advanced optimizations, whereas ASAN$--$ and ESAN are based on older LLVM versions. The time overhead of \sys itself is 5.72\%, which demonstrates the efficiency of \sys. All three variants of \sys also performed admirably. In particular, \sys + PUMM demonstrated the lowest average overhead (6.60\%), surpassing all tools that provide comprehensive heap memory protection. The memory overhead of \sys itself averaged at 54.59\%, and \sys + PUMM also exhibited the best memory overhead among all variants, better than ASAN and ASAN$--$. While it is slightly higher than that of ESAN, we believe it is acceptable given the notable performance improvements it offers. 

To enhance our understanding of the scalability and performance implications of \sys, we measured several parameters related to heap usage. These include the number of instrumentation points, the number of allocation and deallocation operations per second, the amount of shadow memory allocated for each program, and both heap allocation and access frequencies (see Figure \ref{tab:statistic_cpu2017} in the Appendix). These measurements were obtained by inserting counting instructions following each load and store operation, as well as allocation and deallocation functions.  We conducted linear regression to determine how significantly these parameters affect the performance overhead. We found that performance is most closely related to the frequency of heap access. The regression coefficients for other parameters are close to zero, indicating that they have little or no real relationship with time overhead. Based on the frequency of heap accesses, we used 1GHz, 2GHz, and 3GHz as thresholds to categorize the programs into four groups: \textit{lightweight}, \textit{normal}, \textit{heavy}, and \textit{intensive} heap access. We found the geometric mean time overhead of each group is 2.35\%, 3.58\%, 6.52\%, and 29.66\%, respectively. This tiered classification also demonstrates a direct correlation between the intensity of heap access and the performance overhead. Based on the evaluation result, it becomes evident that its adoption is particularly advantageous for lightweight, normal and heavy heap access scenarios, where the method's low to moderate performance overhead can enhance security and monitoring without significantly impacting functionality. On the other hand, for data-intensive and high-performance computing applications, which fall into the intensive heap access categories, the method's higher performance overhead necessitates a more cautious approach.

% Additionally, we noted some observations that may not be intuitive and manually verified their reasons. We noticed that the memory overhead exhibits variability across testcases. In test cases where memory overhead is very low, e.g., \texttt{xz}, this is due to most of the program's memory consumption coming from the stack and \sys doesn't allocate shadow memory for the stack. For testcases with high memory overhead, e.g., \texttt{leela}, the program often allocates small objects. Given \sys's fixed reserved space for each object, if many objects are smaller than this space, the overhead increases significantly. In some tests, e.g., \texttt{parest}, the memory overhead of \sys + FFMalloc exceeds that of \sys alone. It is because the program execute fewer deallocation operations, and \sys's default allocator initially reserves a larger memory pool than FFMalloc, causing \sys to allocate more memory.

Figure \ref{fig:spec2006} presents the time overhead of \sys and DeltaPointer on SPEC2006. \sys exhibits lower overhead in nearly all test cases, with the exceptions being \texttt{h264ref} and \texttt{dealII}. The geometric mean of the time overhead for \sys is also significantly better than that of DeltaPointer. In terms of memory overhead, DeltaPointer achieves an average memory overhead of zero, while \sys incurs an average memory overhead of 68.21\%, which is higher than DeltaPointer. It's worth mentioning that while DeltaPointer may have lower memory overhead, it restricts available address space to 4GB, limiting its scalability for large-scale programs. Additionally, DeltaPointer does not provide defense against underflow errors, whereas \sys offers improved security in this regard. 

Regarding the variation in time overhead observed in SPEC CPU2006 and SPEC CPU2017, which stand at 10.58\% and 5.72\% respectively, we have identified the cause. This discrepancy arises from certain test cases within SPEC2006, such as \texttt{hmmer}, \texttt{dealII}, and \texttt{h264ref}, which contain functions exhibiting distinctive patterns that are challenging for \sys to optimize effectively. To exacerbate matters, the test input often leads to these functions consuming over 50\% of the program's execution time, amplifying the impact of the instrumentation. However, we believe that these issues are more prevalent in smaller-scale programs. Our subsequent experiments demonstrate that larger-scale programs are less likely to encounter such problems.

\paragraph{NGINX} To assess \sys's performance in a real-world, large-scale application, we conducted experiments using Nginx v1.22.1 and the wrk v4.2.0 benchmarking tool. These experiments utilized a configuration of 8 threads, 100 connections, and a test duration of 60 seconds. To ensure consistency, we repeated the test 30 times and recorded the average results. The findings are presented in Table \ref{tab:nginx_eval}. \sys experiences 6.47\% and 2.54\% time overhead for the average and tail (99\%) latencies, respectively; these results illustrate \sys's efficiency and practicality in real-world applications. Among variants of \sys , \sys + MarkUs delivers the best performance, experiencing 27.23\% and 18.98\% time overhead for the average and tail latencies, respectively.

\begin{table}[t]\small
    \tabcolsep=3.2pt
    \centering
    \begin{tabular}{ccccccc}
       \toprule
        \multirow{2}*{\textbf{System}} &
        \multirow{1}*{\textbf{Output}} &
        \multicolumn{5}{c}{\textbf{Latency ($\mu s$)}} \\
        \cline{3-7} \rule{0pt}{11pt} & ($req/s$)
        & \textbf{Average} & \textbf{50\%} & \textbf{75\%} 
        & \textbf{90\%} & \textbf{99\%} \\
       \midrule
        NATIVE          & 158,847 & 611 & 592 & 604 & 623 & 748 \\
        \sys            & 147,550 & 650 & 640 & 649 & 668 & 767 \\         
        SB + MarkUs     & 124,361 & 777 & 759 & 770 & 803 & 890 \\
        SB + FFMalloc   & 110,406 & 870 & 860 & 880 & 900 & 1000 \\ 
        SB + PUMM       & 79,229 & 1220 & 1200 & 1220 & 1270 & 1460 \\ 
        \midrule
    \end{tabular}
    \caption{Evaluation Results of Native, \sys and its variants: Output and Latency Analysis on Nginx. In the Latency column, Average denotes the average latency of the requested connections, while the remaining values depict latency distribution.}
    \label{tab:nginx_eval}
\end{table}

\begin{figure}[t]
    \centering
    \includegraphics[width=\columnwidth]{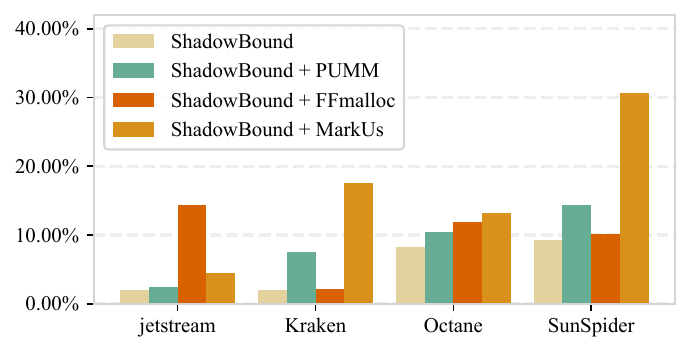}
    \caption{Runtime overhead comparison of \sys and its variants on the Chakra engine: The geometric mean overhead for each system is 4.17\%, 7.28\%, 7.86\%, 13.28\%.}
    \label{fig:chakra-eval}
\end{figure}

\begin{table}[t]\small
    \tabcolsep=3.4pt
    \centering
    \begin{tabular}{cccc}
         \toprule
         \textbf{Website} & \textbf{Native} & \textbf{\sys} & \textbf{Overhead} \\
         \midrule
            www.google.com      & 1202 & 1237 & 2.93\% \\
            www.facebook.com	& 932  & 950  & 2.01\% \\
            www.amazon.com	    & 2399 & 2444 & 1.87\% \\
            www.openai.com	    & 1544 & 1577 & 2.16\% \\
            www.twitter.com	    & 1580 & 1634 & 3.45\% \\
            www.gmail.com	    & 1791 & 1822 & 1.75\% \\
            www.youtube.com	    & 2244 & 2374 & 5.79\% \\
            www.wikipedia.org   & 1085 & 1133 & 4.42\% \\
            www.netflix.com	    & 1415 & 1448 & 2.36\% \\
        \midrule
        Geomean & - & - & 2.74\% \\
        % \bottomrule
    \end{tabular}
    \begin{tabular}{ccccc}
         \toprule
         \textbf{Benchmark} & \textbf{Octane} & \textbf{Kraken} & \textbf{SunSpider} & \textbf{Geomean} \\
         \midrule
         \sys & 3.60\% & 3.30\% & 5.50\% & 4.03\% \\
         \bottomrule
    \end{tabular}
    \caption{Runtime overhead on Chromium: website loading times and JavaScript benchmarks.}
    \label{tab:chromium_eval}
\end{table}

\begin{figure*}[t]
    \centering
    \includegraphics[width=\textwidth]{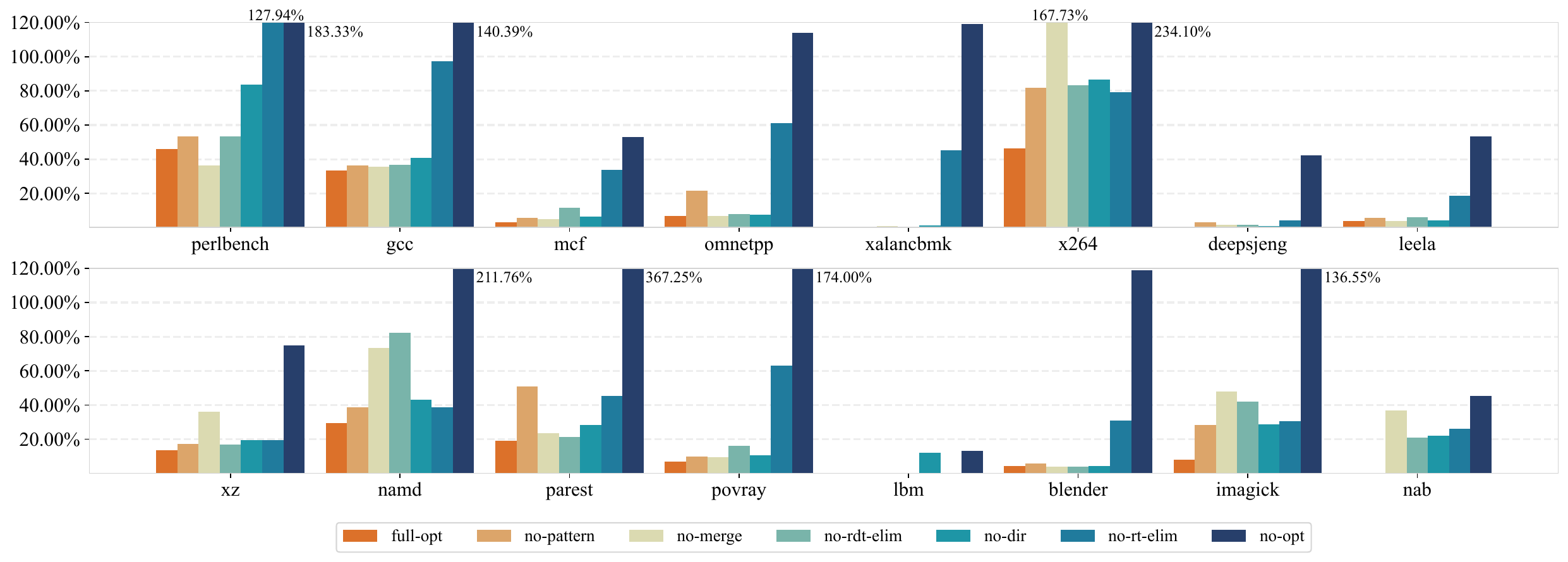}
    \caption{Ablation study result of \sys on SPEC CPU2017. From left to right, the bars show the time overhead of \sys with full optimization, \sys with each optimization disabled, and \sys without optimization. The geomean value of each setting is 5.72\%, 9.51\%, 11.56\%, 11.76\% 12.86\%, 29.28\% and 99.69\%.}
    \label{fig:ablation-study}
\end{figure*}

\paragraph{Chakra} In addition to Nginx, we also evaluated the performance of  \sys, as well as \sys + PUMM, \sys + FFMalloc and \sys + MarkUs on the Chakra. Chakra is a high-performance JavaScript engine developed by Microsoft. Our assessment encompassed four key benchmarks integrated into Chakra: Octane, Kraken, SunSpider, and JetStream. As depicted in Figure \ref{fig:chakra-eval}, our evaluation demonstrates that \sys consistently delivers excellent performance on the Chakra platform. The results indicate that \sys introduces minimal overhead, showcasing its robust compatibility with real-world applications. Particularly noteworthy is the performance of \sys + PUMM and \sys + FFMalloc, which demonstrates exceptional efficiency. These findings underscore \sys's capability to deliver acceptable overheads on large-scale applications.

\paragraph{Chromium} To assess the performance and compatibility of \sys, we conducted an evaluation using the Chromium browser. Chromium is one of the largest and most widely used software, making it an ideal candidate for showcasing \sys's capabilities. We employed three JavaScript benchmarks, Octane, Kraken, and SunSpider, to test \sys's performance. Additionally, we measured the loading times of the 9 most popular websites according to the Top Websites Ranking \cite{top_website}, a critical metric for enhancing the browsing experience. To ensure accurate loading time measurements, we recorded the average loading time for the nine most popular websites.

It is important to note that during our experiments, we encountered difficulties when attempting to dynamically load the allocators for FFMalloc and MarkUs into Chromium; Chromium immediately raised a segmentation fault upon running the program. According to the stack trace, FFMalloc aborted at the re-allocation function, and MarkUs aborted when creating the new thread. Additionally, PUMM requires fuzzing the program and analyzing the trace files. However, PUMM's analysis algorithm proved to be very slow and memory-consuming in our experiments. We attempted analysis for over 24 hours, yet PUMM still failed to produce results and ran out of memory. Consequently, we are presenting results exclusively for the \sys. Our evaluation includes three benchmarks integrated into Chromium and assesses the access speed of nine popular websites. For each benchmark experiment, we conducted 30 repetitions and calculated the geometric mean values to mitigate the impact of random variations. The results are presented in Table \ref{tab:chromium_eval}.Nearly all website loading time overheads remained below 5\%, demonstrating that \sys has minimal influence on the browsing experience for users.

\subsection{Ablation Study}

In our evaluation of \sys's performance on the SPEC CPU2017 suite, we demonstrate the system's performance with full optimization enabled. These optimizations include Runtime-Driven Checking Elimination (rt-elim), Directional Boundary Checking (dir), Security Pattern Identification (pattern), Merge Metadata Operation (merge), and Redundant Checking Elimination (rdt-elim). To isolate the effects of each optimization, we conducted an ablation study consisting of seven experiments. Five of these experiments involved disabling one of these optimizations, while two additional experiments enabled or disabled all optimizations together. These experiments provided valuable insights into the system's complete capabilities and its original performance, allowing us to precisely measure the contributions of each optimization to system performance.

% The results, depicted in Figure \ref{fig:ablation-study}, showcase the substantial performance improvements achieved by \sys's optimization algorithms when compared to the scenario with no optimizations. \sys reduces overhead from 99.69\% to 5.72\%, highlighting the pivotal role these optimizations play. Notably, the Runtime-Driven Checking Elimination shows the best optimization peformance, the overhead increased to 29.28\% when disabling this optimization.
% When disable other optimizations like Directional Boundary Checking Optimization, Merge Metadata Extraction, Redundant Checking Elimination and Security Pattern Identification, the overhead increased to  12.86\%, 11.76\%, 11.56\%, 9.51\% respectively. Even though these optimization performance is not entirely stable, but they are highly useful for certain testcases. The result show the significance of these optimizations in enhancing system performance.

The results presented in Figure \ref{fig:ablation-study} highlight the significant performance improvements achieved with \sys's optimizations, compared to scenarios without them. The optimization reduces overhead from 99.69\% to 5.72\%. Among these optimizations, Runtime-Driven Checking Elimination stands out; without it, the overhead increases to 29.28\%. Disabling other optimizations, specifically Directional Boundary Checking Optimization, Merge Metadata Extraction, Redundant Checking Elimination, and Security Pattern Identification, results in overheads of 12.86\%, 11.76\%, 11.56\%, and 9.51\%, respectively. While the effectiveness of these optimizations may vary across different test cases, each one is essential in the majority of them. These results emphasize the importance of these optimizations in enhancing \sys's performance. Furthermore, the results indicate that moving the checking position from pointer dereference to pointer arithmetic itself does not reduce overhead. As shown in subsection \ref{perf-comp}, the overhead for ASAN is 62.03\%, and the overhead for unoptimized \sys is 99.69\%. The primary difference between them lies in the checking position. Therefore, performing checks at the point of pointer arithmetic does not lead to a reduction in overhead.

\section{Discussion}

\paragraph{Intra-Object Out-Of-Bounds} As \sys checks at the granularity of heap chunks or objects, which is determined at the time of allocation, it does not provide support for preventing out-of-bounds errors that occur within the same object. For instance, it cannot prevent an out-of-bounds access from one array field to another field within the same structure. Consequently, \sys shares the same weakness as prior works \cite{lowfat, duck2018effectivesan, pacmem, kroes2018delta, asan, memcheck}. The boundary within an object's inner fields is not determined by the allocated size but rather by the object's type. If desired, \sys has the capability to relocate shadow memory initialization from the allocation site to the type casting site. We will consider this as a potential area for future work.

\paragraph{Pointer Casting} Given that \sys relies on pointers for boundary checking, it's important to note that developers can write code containing implicit type casts and pointer-integer conversions in low-level languages like C/C++. For implicit type casts, although C/C++ allows this practice, a properly configured LLVM always generates an explicit type casting instruction\cite{opp}. Therefore, \sys can effectively handle implicit type casts. With regard to the conversion between pointers and integers, there is potential for out-of-bounds issues during integer computation instructions. This vulnerability is also noted in several prior works \cite{lowfat, duck2018effectivesan, softbound, PAriCheck}. However, it's crucial to emphasize that developers typically have benign intentions, and some established C/C++ standards \cite{cert_c, misra_c} discourage conversions between integers and pointers unless the developers have a comprehensive understanding of the potential consequences. Consequently, occurrences of such errors are exceptionally rare. In fact, during our extensive evaluation, we did not encounter any such cases. Nevertheless, we acknowledge this as a potential source of false negatives.

\paragraph{Stack Protection} While the design of \sys does not provide protection against stack out-of-bounds vulnerabilities, it is indeed feasible to enhance \sys's capabilities in this regard. LLVM IR employs explicit instructions for allocating objects on the stack. Consequently, \sys can establish the bounds of stack objects by storing this information in the shadow memory at the time of stack allocation. Alternatively, \sys can be integrated with mature solutions such as shadow memory-based solutions \cite{safestack,sok_ss}, hardware-based solutions (e.g., ARM PA \cite{arm_pa}), or hybrid solutions (e.g., Intel CET \cite{intel_ss}). Even though these solutions also use shadow memory, \sys was designed with the possibility of integration in mind, so the shadow memory region of the stack is reserved.
We believe that these methods are sufficient to protect the stack from attacks such as Return-Oriented Programming (ROP) while providing minimal overhead.

% \paragraph{Binary Solution} \sys is not designed for binary targets. However, we believe that \sys might be compatible with certain methods that only work on binaries, such as converting assembly code into IR code, similar to what MEMCHECK\cite{valgrind} does. This would allow \sys to analyze the semantics of the program based on the IR. An extra challenges may arise during instrumentation, especially if it is based on binary patching. This approach might require the insertion of numerous jumps to preserve the original logic of the binary. If the checking instruction is dynamically inserted, the performance could be significantly low.  We will consider
% this as a potential area for future work.

\paragraph{Dynamic Library} \sys is capable of safeguarding a dynamic library, provided it has been instrumented by our tool. Should a dynamic library not undergo instrumentation, \sys retains the ability to detect out-of-bounds errors occurring externally. For instance, if a pointer originating within the library triggers an overflow outside of it, \sys can identify such errors. This is due to \sys's comprehensive monitoring of all memory management APIs, enabling precise tracking of every object's boundaries.

\paragraph{Future Optimization} \sys still possesses optimization potential in both compiler optimization and metadata management aspects. Regarding compiler optimization, \sys does not fully harness the capabilities of Link-Time Optimization (LTO) \cite{lto} and Profile Guided Optimization (PGO) \cite{pgo}. We believe that boundary checking can be further optimized by leveraging additional interprocedural analysis techniques and profiling feedback data. In terms of metadata management, given that \sys is a pure software solution, metadata extraction can be further enhanced by incorporating certain hardware features to maintain the metadata, such as Intel MPX\cite{intel_mpx} and CHERI\cite{cheri_sp}.

% \paragraph{Hardware-assisted protection.}

\section{Related Work}

\paragraph{Checking Optimization} Compiler optimization techniques are commonly employed to optimize out-of-bounds checking. Many previous works \cite{lowfat, duck2018effectivesan, pacmem, sgxbound} have leveraged general optimization techniques to enhance checking performance or eliminate redundant checks. Additionally, there are some works \cite{asan_junxu, zhang2021sanrazor} that have designed custom optimization techniques to optimize Address Sanitizer's \cite{asan} checking. These optimization techniques can reduce a certain amount of overhead but cannot have an order of magnitude impact. The optimization technology of \sys has significantly reduced overhead and has made \sys practical.

\paragraph{Boundary Annotation} LLVM introduced a new feature called \texttt{-fbounds-safety} \cite{fbt}. This feature allows programmers to manually designate length variables for pointers, thereby enabling the compiler to insert boundary checking instructions for these pointers. While a direct performance comparison between \sys and \texttt{-fbounds-safety} is not feasible due to the latter's reliance on human annotations, combining \texttt{-fbounds-safety} with \sys could be beneficial. This is because neither is consistently superior to the other in all scenarios. For pointers with lengths that are already annotated, the \texttt{-fbounds-safety} flag may not require extra instructions to fetch the array length, potentially resulting in better performance. However, \texttt{-fbounds-safety} cannot address out-of-bounds errors caused by type down-casting, while \sys can, as it instruments pointer casting instructions. On the other hand, annotating all pointer length variables requires significant human effort and may not always be feasible. In contrast, \sys can manage these cases automatically without the need for human annotation.

\paragraph{Pointer Tagging} Pointer Tagging is a common technique for storing the boundary in out-of-bounds detection or defense tools. Some studies \cite{kroes2018delta,sgxbound} store complete boundary information in the high-order bits of the pointer, eliminating the need to retrieve boundary information from memory. However, the available high-order bits are not sufficient to contain all the necessary boundary information. These approaches reduce available memory space, which may lead to potential compatibility issues with large-scale programs. To encode more information in the pointers without shrinking the memory space, other studies \cite{duck2018effectivesan,lowfat,TAILCHECK} have implemented custom allocators that allow the original pointer to convey additional boundary information. PACMem \cite{pacmem} uses ARM PA to generate a key stored in the high-order bits. This key is then used as an index to store the actual boundary information in shadow memory. The complexity of these encoding algorithm usually introduce additional arithmetic and memory operations that could potentially impact performance.

\section{Conclusion}

This work introduces \sys, an innovative heap memory protection design. \sys utilizes advanced metadata management and customized compiler optimization to provide a robust heap out-of-bound defense. We integrated it with three state-of-the-art use-after-free defenses without compatibility issues. Ensuring both temporal and spatial memory protection, \sys maintains a minimal performance and memory impact. Experimental results demonstrate its effectiveness and efficiency in defending against heap memory vulnerabilities, especially in programs written in unsafe languages.

\section*{Acknowledgement}

We thank the anonymous reviewers for their valuable suggestions and feedback. This work was partially supported by Northwestern University fellowship, NSF grant CNS-2221122. This paper gives the views of the author, and not necessarily the position of the funding agency.

% abstract + introduction | 1.5  | xxx
% challenges              | 2    | xxxx
% methodology             | 3.5  | xxxxxxx
% implementation          | 1    | xx
% evaluation              | 4    | xxxxxx
% discussion + related    | 1    | x
% total                   | 13   | 12

\bibliographystyle{IEEEtran}
\bibliography{ref}

\appendices
\section{Statistic of SPEC CPU2017 Benchmark}

\begin{minipage}{\textwidth}
    \small
    \centering
            \begin{tabular}{l|cc|cc|c|c|c}
            \toprule
            \multirow{2}{*}{\textbf{Program}} &
              \multicolumn{2}{c|}{\textbf{Instrument Count}} &
              \multicolumn{2}{c|}{\textbf{Operation Per Second}} &
              \textbf{Shadow Memory} &
              \textbf{Alloc Freq} &
              \textbf{Access Freq}
              % \textbf{Time} 
              \\
             &
              \textbf{\#extraction} &
              \textbf{\#checking} &
              \textbf{alloc} &
              \textbf{dealloc} &
              (KB) &
              (MB/s) &
              (GHz) 
              % & (s) 
              \\
            \midrule
            deepsjeng & 9     & 17    & 0.02         & 0.01         & 121160   & 3.59     & 0.06 % & 192 
            \\
            leela     & 113   & 132   & 176,400.76   & 176,400.73   & 58484    & 223.81   & 0.20 % & 305 
            \\
            omnetpp   & 2590  & 3015  & 1,698,303.59 & 1,698,266.30 & 326756   & 214.96   & 0.96 % & 270 
            \\
            \midrule
            nab       & 780   & 1214  & 1,414.20     & 1,077.32     & 176892   & 7.03     & 1.12 % & 238 
            \\
            mcf       & 49    & 99    & 2,224.47     & 2,224.47     & 80364    & 8.58     & 1.23 % & 223 
            \\
            blender   & 15479 & 22445 & 70,776.19    & 70,776.14    & 1297236  & 55.24    & 1.34 % & 163 
            \\
            gcc       & 13571 & 17516 & 816,831.53   & 794,791.98   & 10995204 & 2,935.12 & 1.39 % & 176 
            \\
            xz        & 279   & 631   & 0.45         & 0.37         & 78908    & 13.41    & 1.50 % & 278 
            \\
            lbm       & 10    & 244   & 0.05         & 0.05         & 110196   & 3.17     & 1.85 % & 129 
            \\
            \midrule
            perlbench & 6377  & 7909  & 597,526.88   & 592,088.88   & 657444   & 53.42    & 2.11 % & 268 
            \\
            povray    & 2111  & 2672  & 547.42       & 547.32       & 51372    & 0.09     & 2.45 % & 209 
            \\
            imagick   & 4237  & 5374  & 123,346.30   & 123,346.10   & 61668    & 16.00    & 2.51 % & 302 
            \\
            xalancbmk & 10449 & 12589 & 968,286.71   & 968,286.70   & 647992   & 459.19   & 2.96 % & 143 
            \\
            \midrule
            namd      & 1274  & 4088  & 127.59       & 126.66       & 199664   & 2.69     & 3.26 % & 159 
            \\
            x264      & 4759  & 6677  & 20.68        & 20.67        & 456384   & 1.07     & 3.26 % & 355 
            \\
            parest    & 33368 & 39771 & 631,792.47   & 631,792.30   & 1091476  & 697.83   & 5.34 % & 250 
            \\
            \midrule
            Regr. Coef. &
            $3.88 \times 10^{-4}$ &
            $3.43 \times 10^{-4}$ &
            $2.68 \times 10^{-6}$ &
            $2.53 \times 10^{-6}$ &
            $2.10 \times 10^{-6}$ &
            $6.42 \times 10^{-3}$ &
            4.99 % & /
            \\
            \bottomrule
            \end{tabular}
        \captionof{table}{\textbf{Statistic of SPEC CPU2017 Benchmark}. 
        The table is sorted by \textbf{Access Freq}, heap access frequency, and it is divided into four groups, categorizing programs as lightweight, normal, heavy, and intensive heap access from top to bottom, respectively. The \textbf{Instrument} part indicates the number of instrumentation points inserted by the system. \textbf{Operation Per Second} section indicates the number of allocation and deallocation operations per second. \textbf{Shadow Memory} displays the amount of shadow memory allocated for each program. \textbf{Alloc Freq} represents the rate of memory allocation per second. \textbf{Regr. Coef.}, the linear regression coefficient, indicates the correlation between these indices and performance overhead. The coefficient is close to zero suggesting there is minimal or no relationship between them.
        % \textbf{Time} provides the duration required for each program to complete its execution.
        }
        \label{tab:statistic_cpu2017}
\end{minipage}

\newpage 
$~$
\newpage
\section{Realworld Security Evaluation}

\begin{minipage}{\columnwidth}
    \small
    \centering
        \begin{tabular}{clll}
         \toprule
            \normalsize{\textbf{Source}} &
            \normalsize{\textbf{CVE/Issue ID}} &
            \normalsize{\textbf{Program}} & 
            \normalsize{\textbf{Result}}
            \\
        \midrule
        \multirow{6}{*}{\textsc{SanRazor}}
        & CVE-2015-9101  & lame      &  \ding{52}OD \\ %
        & CVE-2016-10270 & libtiff   &  \ding{52}BR \\ %
        & CVE-2016-10271 & libtiff   &  \ding{52}OD \\ %
        & CVE-2017-7263  & potrace   &  \ding{52}OD \\ %
        & 2017-9167-9173 & autotrace &  \ding{52}OD \\ %
        & 2017-9164-9166 & autotrace &  \ding{52}OD \\ %
        \midrule
        \multirow{5}{*}{\textsc{Asan$--$}} 
        & CVE-2006-6563  & proftpd   &  \ding{52}OD \\ %
        & CVE-2009-2285  & libtiff   &  \ding{52}OD \\ %
        & CVE-2013-4243  & libtiff   &  \ding{52}OD \\ %
        & CVE-2014-1912  & python    &  \ding{52}OD \\ %
        & CVE-2015-8668  & libtiff   &  \ding{52}OD \\ %
        \midrule
        \multirow{6}{*}{\textsc{Magma}} 
        & CVE-2016-1762	 & libxml  &  \ding{52}BR \\ %
        & CVE-2016-1838	 & libxml  &  \ding{52}BR \\ %
        & CVE-2019-10872 & poppler &  \ding{52}OD \\ %
        & CVE-2019-9200	 & poppler &  \ding{52}OD \\ %
        & CVE-2019-7310	 & poppler &  \ding{52}OD \\ %
        & CVE-2013-7443	 & sqlite  &  \ding{52}OD \\ %
        \bottomrule
        \end{tabular}
        \captionof{table}{Security evaluation for \sys on vulnabilities from prior works.}
        \vspace{-4pt}
        \label{tab:prior-realworld-applications}
\end{minipage}

% that's all folks
\end{document}